# Self-propulsion of boiling droplets on thin, heated oil films

Victor Julio Leon and Kripa K. Varanasi
Department of Mechanical Engineering, Massachusetts Institute of Technology,
Cambridge, MA 02139, USA, varanasi@mit.edu

Abstract:
**We report on the self-propulsion of boiling droplets which, despite their contact with viscous, immiscible oil films, attain high velocities comparable to those of levitating Leidenfrost droplets. Experiments and model reveal that droplet propulsion originates from a coupling between seemingly disparate short and long timescale phenomena due to microsecond fluctuations induced by boiling events at the droplet-oil interface. This interplay of phenomena leads to continuous asymmetric vapor release and momentum transfer for high droplet velocities.**

Droplet self-propulsion has been demonstrated using gradients in temperature [1,2], surface tension [3–6], and electric fields [7,8]. Other mechanisms that promote self-propulsion include superhydrophobic asymmetric textures [9–12] and the Leidenfrost effect with asymmetric features or internal flows [13–18]. However, droplet velocities are typically small (~1mm/s), except in cases where droplet friction is minimized (e.g. levitating Leidenfrost droplets, hydrophobic surfaces with electric fields) [7,15–18]. Here we report on the self-propulsion of boiling droplets which, despite their contact with viscous, immiscible oil films, attain high velocities (~10cm/s) comparable to those of levitating Leidenfrost droplets [15–18]. As opposed to previous studies of Leidenfrost droplets on solid surfaces and thick oil pools [19–22], droplets in contact with thin oil films exhibit a unique self-propulsion phenomenon that occurs between the Leidenfrost and boiling regimes (Supplementary Fig. 1). To the best of our knowledge, this self-propulsion phenomenon, resulting from asymmetric vapor ejection mediated by complex transport, wetting, and phase change phenomena, has not been studied before.

On an untextured, plain silicon surface held at 180°C (Fig. 1(a), Supplementary video 1), a DI water droplet initially boils, then levitates in the Leidenfrost state, moving to the right under the influence of gravity. In contrast, Fig. 1(b) (Supplementary video 2) demonstrates how a DI water droplet propels to the right after deposition on a thin, silicone oil film at 180°C. Surfaces are leveled and temperature gradients are minimized (see Supplemental Methods). Astonishingly, droplets in contact with the heated oil films move faster than the levitating Leidenfrost droplets, attaining constant velocities of up to 16 cm/s after heating and accelerating. Reported velocities are extracted by tracking the position of the droplets over time as shown in Fig. 1(c).

FINAL DOI: https://doi.org/10.1103/PhysRevLett.127.074502
PREPRINT

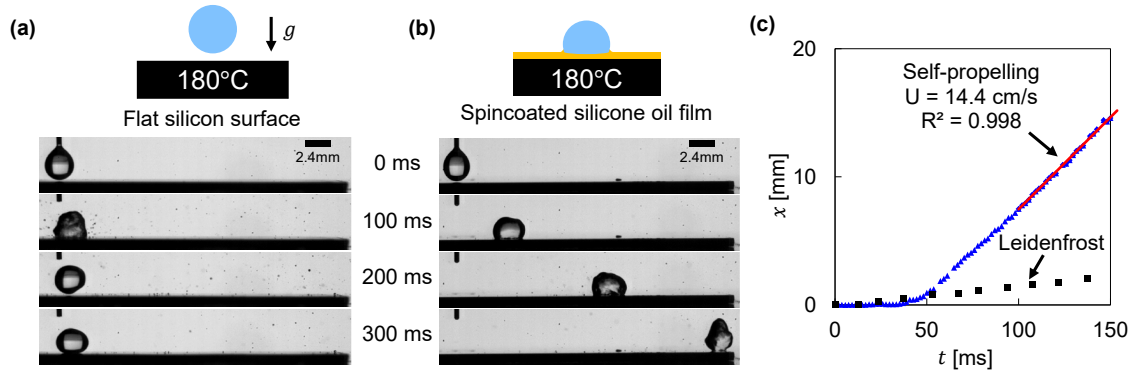

FIG. 1. (a) A sequence of images of a water droplet on an untextured, plain silicon surface at 180°C. The drop first boils then bounces in the Leidenfrost state, moving slowly to the right under the influence of gravity. (b) A water droplet self-propelling on a silicon surface coated with a thin, 13mPas silicone oil film at 180°C. (c) Self-propelling droplet (blue triangles) and Leidenfrost droplet (black squares) position as a function of time for a droplet on a 16mPas silicone oil film at 160°C. Self-propulsion velocities, calculated using the slope of the red line, are constant after droplets heat and accelerate.

Figs. 1(a) and 1(b) demonstrate that, at a given temperature, the inclusion of a silicone oil film suppresses the Leidenfrost effect. Whether the droplet enters the Leidenfrost state or propels appears to be dictated by oil-droplet wetting dynamics. Silicone oil engulfs, or cloaks, water droplets since the spreading coefficients of the oil ($o$) on water ($w$) in air ($a$) ($S_{ow(a)} = \gamma_{wa} - \gamma_{wo} - \gamma_{oa} \sim$ 5mN/m) is positive at elevated temperature [2,23–25]. The velocity at which a thin oil film cloaks a droplet increases as oil viscosity decreases [26,27]. Correspondingly, we observe that as oil viscosity increases, droplets are more likely to lose contact with the oil film as they boil, entering the Leidenfrost state instead of propelling (Supplementary Fig. 2). For experiments conducted with silicone oils of room temperature viscosities from 97 to 970mPas and film temperatures from 100 to 220°C, self-propulsion only occurs consistently between 140 and 180°C. At higher film temperatures, droplets on higher viscosity films tend to enter the Leidenfrost state instead of staying in contact with the oil film and propelling. At lower film temperatures, droplets tend to explode or eject bubbles infrequently instead of propel. In both cases, droplet velocities are far lower than velocities in the self-propelling regime.

Closer inspection of the bottom of a propelling droplet, as shown in a sequence of images in Figs. 2(a), 2(b), and 2(c), provide evidence of asymmetric vapor ejection taking place below a self-propelling droplet. After the initial heating stage [Fig. 2(a)], a vapor bubble nucleates and ejects [Fig. 2(b)] leading to displacement of the droplet in the opposite direction [Fig. 2(c)]. The smooth motion of the drop is a result of continuous ejection of bubbles. In the absence of gravity bias, the nucleation of the first few bubbles is random and appears to dictate the direction of motion. Vapor ejection causes violent disturbances at the oil-water interface, promoting continued bubble nucleation at the same location and continuous asymmetric vapor ejection. There are no specific features on the underlying surface that guide propulsion direction.

The prominence of asymmetric vapor ejection observed in experiments suggests that the force driving the droplet motion is the rate of momentum change due to vapor ejection:

$$\frac{dp}{dt} = \frac{d(mV)}{dt} = \dot{m}V + m\dot{V} \tag{1}$$

where $p$ is momentum, $V$ is the vapor ejection velocity, and $\dot{m}$ is the water vaporization rate. We emphasize that $V$, the vapor ejection velocity, is not the droplet ejection velocity (see Supplementary discussion on droplet ejection). An estimate for $V$ is determined experimentally using the Bernoulli equation ($\Delta P_{ba} = \rho V^2/2$), where $\Delta P_{ba}$ is the difference between the pressure inside the bubble ($b$) immediately before ejection and atmospheric ($a$) pressure. $\Delta P_{ba}$ is calculated using the Young-Laplace equation ($\Delta P = 2\gamma_{wa}/R_i$), which relates the pressure rise across a spherical interface, $\Delta P$, to the interface's radius of curvature, $R_i$. $\Delta P_{ba}$ is calculated as the sum of $\Delta P$ for the atmosphere-droplet and droplet-bubble interface. An image prior to a typical bubble ejection is shown in Fig. 2(b) from which the droplet and ejecting bubble radii are measured. Bubbles are observed to eject at radii between 0.5 and 1mm, which correspond to $V$ ranging from 24 to 29m/s. At such gas ejection velocities, ligaments are expected to form via shear stripping during vapor ejection [28], which we observe beneath propelling droplets (Fig. 2(c), Supplementary video 3). The ligament velocities we see (~10cm/s) also match literature values [28] for gas velocities of ~25m/s.

As ejecting bubble radii, and consequently $V$, vary little in the self-propulsion regime, we neglect the $\dot{V}$ term in Equation 1:

$$\frac{dp}{dt} \sim \dot{m}V \tag{2}$$

In order to determine $\dot{m}$, we consider the energy balance in an oil film with dynamic viscosity $\mu$, film thickness $h$, density $\rho$, thermal conductivity $k$, and specific heat $c_p$ and a droplet of radius $R$ propelling at velocity $U$ with oil contact area $A$, composed of a liquid with a heat of vaporization $h_{fg}$. A small Brinkman number ($Br = \mu U^2/k/\Delta T \sim 10^{-4}$) indicates that viscous dissipation in the thin oil film is negligible compared to the total heat conduction through the film ($\Delta T = T_s - T_{sat}$, where $T_s$ is the substrate temperature and $T_{sat}$ is the saturation temperature of water at atmospheric pressure, 100°C). Therefore, the vaporization rate in the system is set by the heat conducted from the oil film to the bottom of the water droplet at the oil-water interface. The energy balance per unit time is:

$$\dot{m}h_{fg} \sim k\frac{dT}{dz}A \sim k\frac{dT}{dz}\pi R^2 \tag{3}$$

where $z$ indicates the vertical direction, as shown in Fig. 2(a).

The change in momentum due to the ejecting vapor is balanced by the viscous shear force in the oil wetting ridge at the circumference of the droplet base [23,29]. This force is the result of the Landau-Levich entrainment of oil under the droplet as the droplet moves [29]. Accordingly, the drag force scales with the capillary number ($Ca = \mu U/\gamma_{oa}$). Because $\mu_w \ll \mu_o$, the viscous shear force in the droplet and the viscous shear force in the rest of the oil-water interfacial area ($F_{ow} \sim \mu_w U R$) are negligible compared to the Landau-Levich force. The viscous shear force $F_f$ at the outer circumference of the droplet base then scales as [29]:

$$F_f \sim 2\pi l R \left(\mu \frac{dU}{dz}\right) \tag{4}$$

where $l \sim RCa^{1/3}$ is the length scale of the deformation of the droplet at the circumference of the droplet base. Substituting for $l$:

$$F_f \sim 2\pi R^2 \left(\mu \frac{dU}{dz}\right) Ca^{1/3} \qquad (5)$$

Combining Equations 2, 3, and 5 yields:

$$\left(k \frac{dT}{dz} \frac{1}{h_{fg}}\right) V \sim 2\mu \frac{dU}{dz} Ca^{1/3} \qquad (6)$$

Interestingly, our scaling predicts that $U$ has no dependence on $R$, which is confirmed by experiments [Fig. 2(d)]. We limit our study below R~1mm, since at larger droplet sizes, bubbles begin to rise into the body of the puddles instead of ejecting out the side to cause self-propulsion (see Supplementary video 4 and discussion in Supplementary information).

$\Delta T$, $U$, and $h$ are taken as the characteristic scales of the temperature and velocity gradients. Substituting $dT/dz \sim \Delta T/h$ and $dU/dz \sim U/h$ into Equation 6 produces a $U \sim \mu^{-1}$ scaling which does not agree with experiments varying oil viscosity [Fig. 2(e)].

We hypothesize that bubbling and vapor ejection due to boiling at the oil-water interface disrupts the momentum and thermal boundary layer beneath the droplet. Accordingly, the temperature and velocity gradient length scales in the oil film are estimated using characteristic thermal and momentum boundary length scales $\delta_T = \sqrt{\alpha \tau_T}$ and $\delta_U = \sqrt{\nu \tau_U}$ instead of $h$:

$$U \sim \frac{1}{2} \sqrt{\frac{kc_p}{\mu} \frac{\tau_U}{\tau_T}} \frac{\Delta T}{h_{fg}} V \left(\frac{\mu U}{\gamma_{oa}}\right)^{-1/3} \qquad (7)$$

$$U \sim \frac{1}{\mu^{5/8}} \left(\frac{1}{2} \sqrt{kc_p \frac{\tau_U}{\tau_T} \frac{\Delta T}{h_{fg}}} V \gamma_{oa}^{1/3}\right)^{3/4} \qquad (8)$$

where $\alpha = k/\rho c_p$ is the thermal diffusion coefficient, $\nu = \mu/\rho$ is the momentum diffusion coefficient, and $\tau_U$ and $\tau_T$ are characteristic momentum and thermal time scales, respectively. The $U \sim \mu^{-5/8}$ scaling agrees well with experiments [Fig. 2(e)].

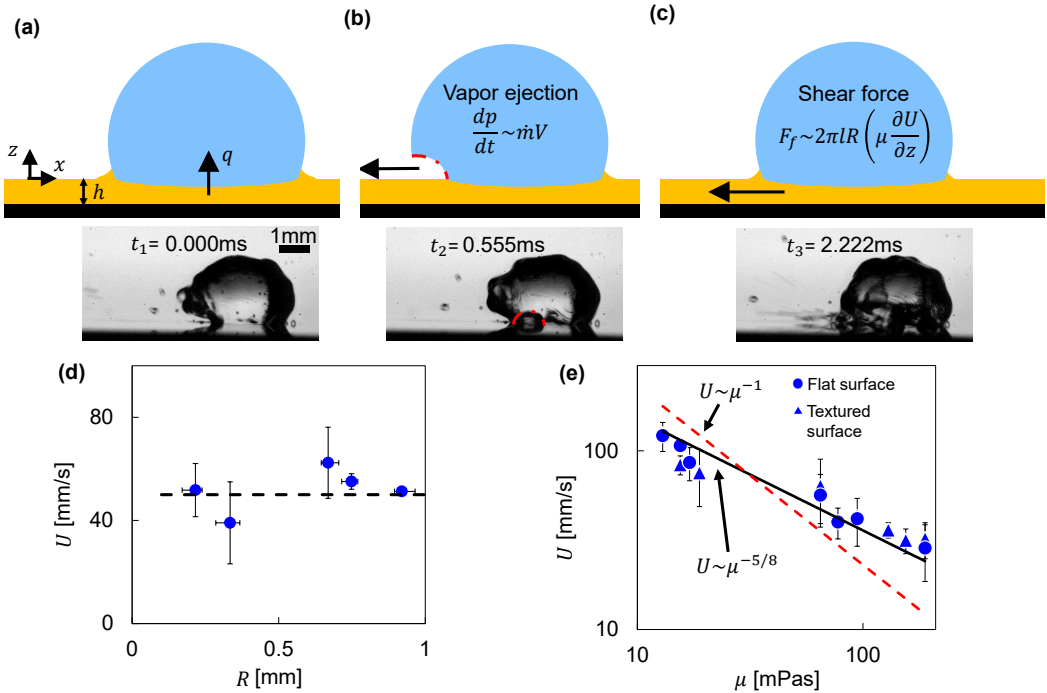

FIG. 2. The images below each schematic show frames from a high speed video of a self-propelling droplet ejecting a bubble on a 130mPas oil film at 180°C. (a) Heat transfer from the heated oil film to the droplet drives the evaporation of the droplet. (b) A bubble grows and is ejected, propelling the droplet to the right. The dashed red line denotes the outline of an ejecting bubble. (c) Filaments form due to shear stripping during vapor ejection. Momentum ejection is balanced by the viscous shear force at the circumference of the droplet's base. (d) Droplet velocity is independent of droplet radius, as predicted by the scaling model. The dashed black line denotes the average velocity of 50mm/s. Experiments were conducted on 19mPas oil films at 140°C. Horizontal error bars denote the range of droplet radii included in each point. Vertical error bars correspond to the standard deviations. (e) Droplet velocities as a function of oil viscosities at 140, 160, and 180°C for flat and textured surfaces (see Supplemental Methods). Droplet velocities are not significantly affected by surface texture. The solid black and dashed red line fits correspond to $U = 8.5\mu^{-5/8}$ and $U = 2.3\mu^{-1}$, respectively. Reported velocities are the average of 3-6 self-propulsion events.

As noted before, the scaling model relies on the hypothesis that the thermal and momentum boundary layers in the oil film are not fully developed under the droplet. This means that $U$ should be independent of oil film thickness, $h$. The film thickness beneath a propelling droplet varies with $U$ and $\mu$, since $h \sim R(\mu U)^{2/3}$ as a result of Landau-Levich entrainment [29]. For our experimental range of $R$, $U$, and $\mu$, entrainment leads to $h$ in the range of 100 to 500 $\mu$m beneath the propelling droplets. Additionally, for a given temperature and oil viscosity, $U$ does not vary significantly between flat and textured surfaces (see Supplemental Methods), implying that $U$ is independent of $h$.

To further validate quantitatively that the dissipations are confined to their respective boundary layers, $\delta$, we compare the characteristic timescale for momentum diffusion ($\tau_U \sim h^2/\nu$) and thermal diffusion ($\tau_T \sim h^2/\alpha$). At time scales greater than $\tau$, the respective boundary layer is fully developed and dissipations are confined by $h$, not $\delta$. For our experiments, $h \sim 100 \mu m$, so $\tau_U \sim 100 \mu s$ and $\tau_T \sim 100 ms$.

Notably, the momentum condition is more restrictive than the thermal condition. To observe phenomena occurring on the rapid timescales suggested by the momentum scaling, we take 100,000fps high speed videos of self-propelling droplets. Surface fluctuations are observed to occur on $10 \mu s$ timescales near the base of self-propelling droplets during and between vapor ejection events (Fig. 3(b), Supplementary video 3 and 5). In contrast, surface fluctuations are not observed in the Leidenfrost droplet control case (Fig. 3(c), Supplementary video 5), implying that boiling causes disturbances at $10 \mu s$ timescales below the boiling droplets. Fig. 3(d) shows normalized image intensity over time created from the same videos, quantitatively reinforcing the surface fluctuation frequencies.

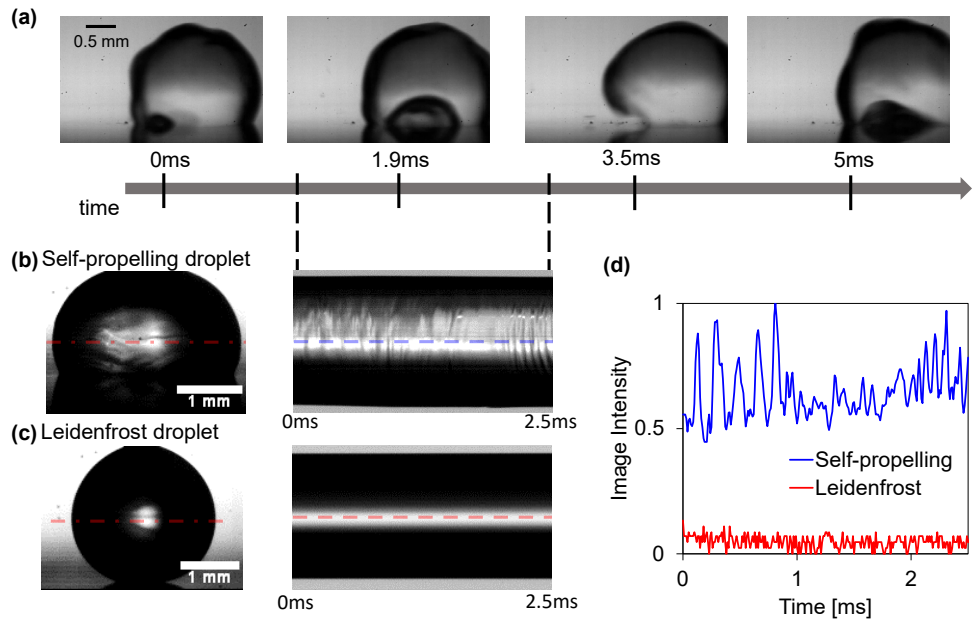

FIG. 3. The proposed model and experimental evidence suggest that the droplet velocity is a result of boiling phenomena occurring on two significantly different timescales. (a) Bubble ejections occurring at millisecond timescales result in the overall momentum ejection and propulsion velocities (Supplementary video 6). The images are of a droplet propelling from left to right on a 13mPas silicone oil film at 180°C. (b,c,d) Evidence of boiling events occurring on the $10 \mu s$ timescale under the droplet that can affect the thermal and momentum boundary layers in the oil film, producing the $U \sim \mu^{-5/8}$ observed in Fig. 2(e) and explained by the model. Full high speed videos are included as Supplementary video 5. (b,c), Spatiotemporal diagrams on the right are created using the line of pixels on the dotted-dashed red lines on the images on the left. (b) Self-propelling droplet demonstrating rapid surface fluctuations. Videos were taken on a 19mPas oil film at 140°C between vapor ejection events. (c) In contrast, a Leidenfrost droplet has no notable surface fluctuations. (d) Normalized image intensity over time along the dashed lines in the spatiotemporal diagrams. The timescale of the fluctuations are on the order of $10 \mu s$.

We propose that the observed disturbances disrupt the normal development of the momentum and thermal boundary layers in the oil film and dictate $\tau_U$ and $\tau_T$ in Equation

8. The surface fluctuations indicate that the droplet is vibrating on the oil film at $10\mu s$ timescales, distorting the momentum boundary layer during propulsion. Moreover, the fluctuations appear to result from bubbling events that occur at the oil-droplet interface in multiple locations (Supplementary video 3 and 5). The vapor bubbles are less thermally conductive than water (0.6 vs 0.03 W/m/K, respectively), modifying the thermal boundary condition at the oil-droplet interface and disrupting the thermal boundary layer at the same $10\mu s$ timescale. Thus, $\tau_U$ and $\tau_T$ are of similar order (~$10\mu s$), canceling each other in Equation 8.

To further validate the model, Equation 8 is rearranged into dimensionless form,

$$\frac{U}{V} \sim \frac{1}{2}\left(\frac{Ja}{Pr^{1/2}Ca^{1/3}}\right) \quad (9)$$

where $V$ is the average vapor ejection velocity (~26m/s), the Jakob number ($Ja = c_p\Delta T/h_{fg}$) is the ratio of sensible heat in oil to the latent heat of water, and the Prandtl number ($Pr = \nu/\alpha$) is the ratio of the momentum diffusivity and thermal diffusivity. All experiments with varied surface texture, viscosities in the range of 10 to 200mPas, and temperatures in the range of 140 to 180°C are plotted in Fig. 4 according to the relations predicted by Equation 9. A prefactor of 0.9 is found with $R^2$=0.87, demonstrating that the model captures the relevant physics of the self-propulsion mechanism.

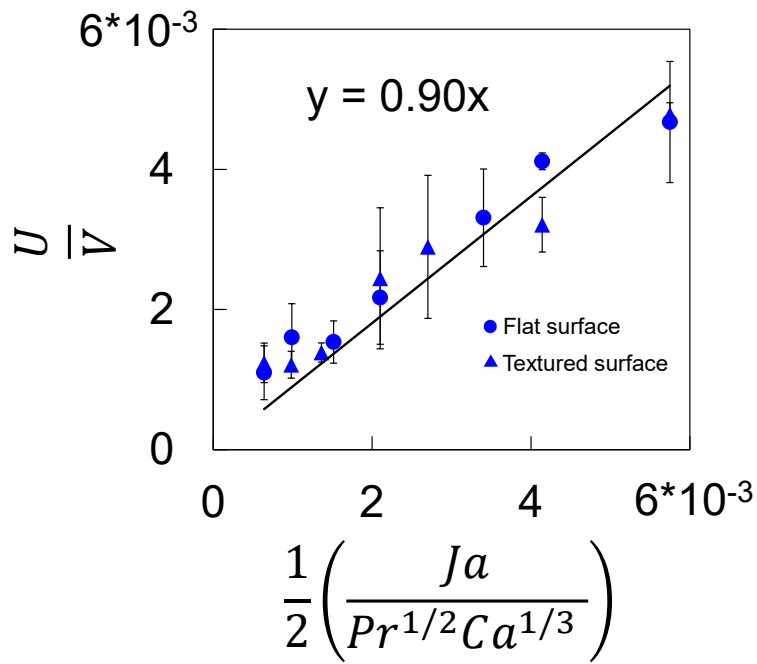

FIG. 4. Dimensionless velocity plotted against the dimensionless parameter predicted by the scaling law in Equation 9. Experiments, which vary viscosity from 10 to 200mPas, temperature from 140 to 180°C, and surface texture, collapse onto a line with slope 0.9 and $R^2$=0.87. Reported velocities are the average of 3-6 self-propulsion events, and error bars correspond to standard deviations.

The self-propulsion mechanism investigated here demonstrates that, counterintuitively, a droplet in contact with a viscous film can reach velocities comparable to those of levitating Leidenfrost droplets on millimetric ratchets. A scaling law, in agreement with experimental droplet velocities, reveals that droplet velocities are a product of the significant coupling of seemingly disparate timescale phenomena due to boiling phenomena occurring at microsecond timescales. The model has been validated experimentally across a range of temperatures, viscosities, droplet radii, and surface textures. These insights could be broadly applied to other vapor release systems such as sublimating solids and volatile, reactive and fizzy droplets. Further studies on the formation of the initial asymmetry and the rich oil-droplet cloaking dynamics may shed light on how to control the Leidenfrost point on thin liquid films and the self-propulsion direction on a surface. Such a surface could quickly and controllably shed corrosive and fouling droplets from heated surfaces.

## Acknowledgements

This material is based upon work supported by the National Science Foundation Graduate Research Fellowship under Grant No. 1122374. We thank M. Costalonga, A. Moreno Soto, V. Jayaprakash, S. Panat, J. Lake, H. Kang, H.-L. Girard, S. Mowlavi, B. Blanc, and P.J. Pow-Sang for insightful discussions and thank M. Costalonga, A. Moreno Soto, V. Jayaprakash, B. Blanc, and S. Dhulipala for proof-reading the manuscript.

## Author contributions

V.J.L. and K.K.V. designed the experiments, analyzed the data, developed the models, and wrote the manuscript. V.J.L. carried out the experiments.

## References

[1] Y. Zhao, F. Liu, and C.-H. Chen, *Thermocapillary Actuation of Binary Drops on Solid Surfaces*, Appl. Phys. Lett. **99**, 104101 (2011).

[2] N. Bjelobrk, H.-L. Girard, S. Bengaluru Subramanyam, H.-M. Kwon, D. Quéré, and K. K. Varanasi, *Thermocapillary Motion on Lubricant-Impregnated Surfaces*, Phys. Rev. Fluids **1**, 63902 (2016).

[3] E. Bormashenko, Y. Bormashenko, R. Grynyov, H. Aharoni, G. Whyman, and B. P. Binks, *Self-Propulsion of Liquid Marbles: Leidenfrost-like Levitation Driven by Marangoni Flow*, J. Phys. Chem. C **119**, 9910 (2015).

[4] Z. Izri, M. N. van der Linden, S. Michelin, and O. Dauchot, *Self-Propulsion of Pure Water Droplets by Spontaneous Marangoni-Stress-Driven Motion*, Phys. Rev. Lett. **113**, 248302 (2014).

[5] P. E. Galy, S. Rudiuk, M. Morel, and D. Baigl, *Self-Propelled Water Drops on Bare Glass Substrates in Air: Fast, Controllable and Easy Transport Powered by Surfactants*, Langmuir (2020).

[6] C. H. Meredith, P. G. Moerman, J. Groenewold, Y.-J. Chiu, W. K. Kegel, A. van Blaaderen, and L. D. Zarzar, *Predator–Prey Interactions between Droplets Driven by Non-Reciprocal Oil Exchange*, Nat. Chem. **12**, 1136 (2020).

[7] M. Gunji and M. Washizu, *Self-Propulsion of a Water Droplet in an Electric Field*, J. Phys. D. Appl. Phys. **38**, 2417 (2005).

[8] Q. Sun, D. Wang, Y. Li, J. Zhang, S. Ye, J. Cui, L. Chen, Z. Wang, H.-J. Butt, D. Vollmer, and X. Deng, *Surface Charge Printing for Programmed Droplet Transport*, Nat. Mater. **18**, 936 (2019).

[9] T. M. Schutzius, S. Jung, T. Maitra, G. Graeber, M. Köhme, and D. Poulikakos, *Spontaneous Droplet Trampolining on Rigid Superhydrophobic Surfaces*, Nature **527**, 82 (2015).

[10] K. M. Wisdom, J. A. Watson, X. Qu, F. Liu, G. S. Watson, and C.-H. Chen, *Self-Cleaning of Superhydrophobic Surfaces by Self-Propelled Jumping Condensate*, Proc. Natl. Acad. Sci. U. S. A. **110**, 7992 (2013).

[11] P. Papadopoulos, L. Mammen, X. Deng, D. Vollmer, and H.-J. Butt, *How Superhydrophobicity Breaks Down*, Proc Natl Acad Sci USA **110**, 3254 (2013).

[12] C. Stamatopoulos, A. Milionis, N. Ackerl, M. Donati, P. Leudet de la Vallée, P. Rudolf von Rohr, and D. Poulikakos, *Droplet Self-Propulsion on Superhydrophobic Microtracks*, ACS Nano (2020).

[13] A. Bouillant, T. Mouterde, P. Bourrianne, A. Lagarde, C. Clanet, and D. Quéré, *Leidenfrost Wheels*, Nat. Phys. **14**, 1188 (2018).

[14] G. Launay, M. S. Sadullah, G. McHale, R. Ledesma-Aguilar, H. Kusumaatmaja, and G. G. Wells, *Self-Propelled Droplet Transport on Shaped-Liquid Surfaces*, Sci. Rep. **10**, 14987 (2020).

[15] H. Linke, B. J. Alemán, L. D. Melling, M. J. Taormina, M. J. Francis, C. C. Dow-Hygelund, V. Narayanan, R. P. Taylor, and A. Stout, *Self-Propelled Leidenfrost Droplets*, Phys. Rev. Lett. **96**, 154502 (2006).

[16] G. Lagubeau, M. Merrer, C. Clanet, and D. Quere, *Leidenfrost on a Ratchet*, Nat. Phys. **7**, 395 (2011).

[17] A. Gauthier, C. Diddens, R. Proville, D. Lohse, and D. van der Meer, *Self-Propulsion of Inverse Leidenfrost Drops on a Cryogenic Bath*, Proc. Natl. Acad. Sci. **116**, 1174 (2019).

[18] J. T. Ok, E. Lopez-Oña, D. E. Nikitopoulos, H. Wong, and S. Park, *Propulsion of Droplets on Micro- and Sub-Micron Ratchet Surfaces in the Leidenfrost Temperature Regime*, Microfluid. Nanofluidics **10**, 1045 (2011).

[19] L. Maquet, B. Sobac, B. Darbois-Texier, A. Duchesne, M. Brandenbourger, A. Rednikov, P. Colinet, and S. Dorbolo, *Leidenfrost Drops on a Heated Liquid Pool*, Phys. Rev. Fluids **1**, 53902 (2016).

[20] B. Sobac, L. Maquet, A. Duchesne, H. Machrafi, A. Rednikov, P. Dauby, P. Colinet, and S. Dorbolo, *Self-Induced Flows Enhance the Levitation of Leidenfrost Drops on Liquid Baths*, Phys. Rev. Fluids **5**, 62701 (2020).

[21] A.-L. Biance, C. Clanet, and D. Quéré, *Leidenfrost Drops*, Phys. Fluids **15**, 1632 (2003).

[22] D. Quéré, *Leidenfrost Dynamics*, Annu. Rev. Fluid Mech. **45**, 197 (2013).

[23] J. D. Smith, R. Dhiman, S. Anand, E. Reza-Garduno, R. E. Cohen, G. H. McKinley, and K. K. Varanasi, *Droplet Mobility on Lubricant-Impregnated Surfaces*, Soft Matter **9**, 1772 (2013).

[24] Y. H. Mori, N. Tsui, and M. Kiyomiya, *Surface and Interfacial Tensions and Their Combined Properties in Seven Binary, Immiscible Liquid-Liquid-Vapor Systems*, J. Chem. Eng. Data **29**, 407 (1984).

[25] E. Ricci, R. Sangiorgi, and A. Passerone, *Density and Surface Tension of Dioctylphthalate, Silicone Oil and Their Solutions*, Surf. Coatings Technol. **28**, 215 (1986).

[26] A. Carlson, P. Kim, G. Amberg, and H. A. Stone, *Short and Long Time Drop*

*Dynamics on Lubricated Substrates*, Europhys. Lett. **104**, 34008 (2013).

[27] A. A. Günay, S. Sett, Q. Ge, T. Zhang, and N. Miljkovic, *Cloaking Dynamics on Lubricant-Infused Surfaces*, Adv. Mater. Interfaces (2020).

[28] P. Marmottant and E. Villermaux, *On Spray Formation*, J. Fluid Mech. **498**, 73 (2004).

[29] A. Keiser, P. Baumli, D. Vollmer, and D. Quéré, *Universality of Friction Laws on Liquid-Infused Materials*, Phys. Rev. Fluids **5**, 14005 (2020).

[30] See Supplemental Material, which includes Refs. [31-34], for supplementary figures, supplementary methods, and discussions on droplet ejection statistics and maximum droplet radii for self-propulsion.

[31] D. Soto, H.-L. Girard, A. Le Helloco, T. Binder, D. Quéré, and K. K. Varanasi, *Droplet Fragmentation Using a Mesh*, Phys. Rev. Fluids **3**, 83602 (2018).

[32] A. Eifert, D. Paulssen, S. N. Varanakkottu, T. Baier, and S. Hardt, *Simple Fabrication of Robust Water-Repellent Surfaces with Low Contact-Angle Hysteresis Based on Impregnation*, Adv. Mater. Interfaces **1**, 1300138 (2014).

[33] J. H. Koschwanez, R. H. Carlson, and D. R. Meldrum, *Thin PDMS Films Using Long Spin Times or Tert-Butyl Alcohol as a Solvent*, PLoS One **4**, e4572 (2009).

[34] Clearco, *Viscosity to Temperature Chart*, http://www.clearcoproducts.com/pdf/bath-fluids/silicone_fluids_viscosity_temp_chart.pdf.